\documentclass[preprint]{elsarticle}
\usepackage{graphicx,subfigure}
\usepackage{amsmath,amssymb}
\usepackage{hyperref}
\usepackage{hhline}
\usepackage{braket}
\usepackage{bm}

\newcommand{\be}{\begin{eqnarray}}
\newcommand{\ee}{\end{eqnarray}}

\def\slashchar#1{\setbox0=\hbox{$#1$}           
   \dimen0=\wd0                                 
   \setbox1=\hbox{/} \dimen1=\wd1               
  \ifdim\dimen0>\dimen1                        
 \rlap{\hbox to \dimen0{\hfil/\hfil}}      
  #1                                        
 \else                                        
    \rlap{\hbox to \dimen1{\hfil$#1$\hfil}}   
    /                                         
 \fi}                                         %

\begin{document} 

\title{Nucleon clustering at kinetic freezeout of heavy-ion \\
collisions via path-integral Monte Carlo}

\author{Dallas DeMartini and Edward Shuryak}

\address{Center for Nuclear Theory, Department of Physics and Astronomy, \\ Stony Brook University,
Stony Brook NY 11794-3800, USA}

\begin{abstract}
	Clustering of the four-nucleon system at kinetic freezeout conditions is studied using path-integral Monte Carlo techniques. This method seeks to improve upon previous calculations which relied on approximate semiclassical methods or few-body quantum mechanics. 
	Estimates are given for the decay probabilities of the 4N system into various light nuclei decay channels and the strength of spatial correlations is characterized. Additionally, a simple model is presented to describe the impact of this clustering on nucleon multiplicity distributions. The effects of a possible modification of the inter-nucleon interaction due to the 
	close critical line (and hypothetical QCD critical point) on the clustering are also studied. 
\end{abstract}


\maketitle

\section{Introduction}

  There has been a renewed interest in heavy-ion collision experiments at energies $\sqrt{s_{NN}}$
  of the order of few GeV.
 The most prominent of current ones are the Beam Energy Scan (BES) at the Relativistic Heavy-Ion Collider (RHIC) and at HADES experiment at GSI. New collider facilities
 are under construction at GSI, Darmstadt and JINR, Dubna, with
 planned fixed-target facilities in a few other laboratories. Baryon-rich matter produced
 there is much less understood that QGP produced at LHC and higher
energies of RHIC. 
  Stage I of BES program studied Au + Au collisions at energies $\sqrt{s_{NN}} =$ 5 - 39 GeV,
  with stage II going to lower energies and fixed target inside the STAR detector. 
  Among its goals is the search for signals of the hypothetical QCD critical point.
  One set of observable watched in it are distributions of baryons and their cumulants  \cite{Thader:2016gpa}, another
 is light nuclei production \cite{Liu:2019ppd}. Both will be discussed in our paper. 
 
Hadronic composition is known to be well described by thermal conditions at chemical freezeout, which by now are well mapped at the phase diagram for LHC and higher
energies of RHIC. An important
component of it is correct account of ``feed-down" from hadronic resonances. However
at higher values of baryon chemical potential $\mu$ (lower collision energies) one
finds more copious production of light nuclei, and for those various  ``feed-downs" 
from intermediate states and resonances are not yet carefully studied.
Simple nucleon coalescence models \cite{Sun:2018jhg} suggest that the production ratios for light nuclei (such as the $N_tN_p/N_d^2$ ratio we will use) should be independent of beam energy, and being close in value to the ratio of statistical weights. The current data, admittedly
not yet very precise, does not support these statements and suggests 
interesting energy dependence. 

Spatial correlations of the nucleons at freezeout -- the subject of this paper --
are supposed to explain these phenomena in detail.
Previous works have elucidated a special role of the 4-nucleon system, which 
is the smallest one possessing multiple near-zero bound states and resonances.  
It is also robust enough to withstand
the relevant temperatures at kinetic freezeout $T_{kin}\sim 110$ MeV.
Recent work has suggested that modifications to the production of light nuclei may be a signal of correlations and fluctuations induced by critical phenomena \cite{Oliinychenko:2020ply}.

The  thermal/quantum density matrix of several nucleons has been evaluated
in Ref. \cite{Shuryak:2019ikv} using two approximations:
(i) a novel semiclassical method based on ``flucton" solution;  and (ii) the solution of hyperradial
Schrodinger equation in $3(N-1)$ dimensions.
 The latter  explicitly found the second ``hyperspherical" $J^{P}=0^{+}$ bound state, one of multiple excited states of $^4 He$, with relevant  two-body decay channels $d+d$, $t+p$, or $^3He+n$. But, since among the experimentally known resonances one finds  many
states with orbital momentum $L=1$ and   $L=2$, one would need more general
method to study ``preclustering" in this system.  
 
In this paper we 
treat thermal/quantum dynamics of few-nucleon systems from first principles, using the most straightforward numerical method, the path-integral Monte Carlo (PIMC).
We also focus on  alpha-particle, for three reasons. First of all, it is the largest $N=4$
corresponding to the number of different nucleon spin-isospin states. Therefore all four nucleons  $p^\uparrow p^\downarrow n^\uparrow n^\downarrow$  are distinguishable,
which greatly simplifies PIMC application. 
(It is for this reason that PIMC was first used for nuclear systems by one of us long ago \cite{Shuryak:1984xr}.)

The second reason: it is the first system  to withstand the relevant temperatures. Let us remind the reader that the binding energy of $^4 He$ is $E_B=$ 28.3 MeV, while the binding energy of $d$ is only $E_B=$ 2.2 MeV. For ``preclusters", which are statistical correlations, the temperature should be compared to maximal value of the
potentials holding a nucleon, $(N-1)V(r)\sim 150\, MeV$.
The third reason we already mentioned: unlike lighter $d,t$ systems, it has many excited states.

Ultra-relativistic heavy-ion collisions produce hot, deconfined QCD matter, the Quark-Gluon Plasma (QGP). The QGP formed in these collisions quickly expands and cools. As it cools, the quarks and gluons become re-confined and hadronize. Two other transitions
happen on the way, known as the chemical and kinetic freezeouts. 
The former one occurs very close to phase transitions line, it is marked by the fact that
when the system has cooled enough for inelastic collisions between hadrons to cease. At this point, the total yields of all produced particles are fixed. The temperature $T_{ch}$ and baryon chemical potential $\mu^B_{ch}$ of the are found by use of a statistical hadronization model such as \cite{Andronic:2005yp}, which fits thermal parameters $T_{ch}$ and $\mu_{ch}$ from particle yield ratios.
For some time after that, elastic collisions continue until  the system has expanded so much that elastic collisions cease as well, at the $kinetic$ freezeout. Because this is the end of interactions for these particles, the transverse momentum $p_T$ is fixed beyond this time. Thermal values
$T_{kin}$ and $\mu_{kin}$  of the kinetic freezeout are calculated from 
elastic collision rates, usually using hydrodynamics \cite{Teaney:2000cw}
or simplified "blast wave'' models \cite{Schnedermann:1993ws}. 
See Table \ref{tab_FO} for the specific values for relevant collision energies.  

This paper is laid out as follows: In Section II, we discuss the basic details of the Monte Carlo simulation and some preliminary calculations. Section III lays out the idea of the nucleon 'precluster' and results of simulations of the kinetic freezeout conditions. Section IV looks at the effects of modified inter-nucleon interactions and Section V describes the observables that can be estimated from this work. 

If there exists the QCD critical point, a near-massless critical mode should be strongly fluctuating in its vicinity. 
These fluctuations are very nonlinear, inducing certain 3- and 4-nucleon forces. Calculations of those forces and their effect on nucleon clustering are not discussed below and are to be found in our subsequent paper Ref. \cite{DeMartini:2020anq}.

\section{The setting}
\subsection{The path integral at finite temperatures and nucleon interactions} 

Feynman's formulation of quantum statistical mechanics is based on the path integral:
\begin{equation}
\bra{\vec{x}(t_f)}e^{-\hat{H}(t_f - t_i)}\ket{\vec{x}(t_i)} = \int Dx(t) e^{-S[\vec{x}]},
\end{equation}
where $\int Dx(t)$ is the integral over all periodic paths $\vec{x}(t+\beta) = \vec{x}(t)$ with a period $\beta=\hbar/T$ 
in Euclidean time. 
Numerical evaluation of this path integral uses discretized time, into $N_t$ time slices with spacing $a= \beta/N_t$. Depending on the temperature discussed, we use $N_t$ of hundreds or even thousands.

The discretized Euclidean action is
\begin{equation}
S[\vec{x}] = a\sum_{j=1}^{N_t} \left(\frac{m}{2a^2} (\vec{x}_{j+1} - \vec{x}_j)^2 + V(\vec{x_j}) \right).
\label{eq_act}
\end{equation}
Here $\vec{x}_j$ is the position of the particle on the j-th time slice and the periodicity condition is imposed by setting $\vec{x}_{N_t + 1} = \vec{x}_1$. This discretized action is exact in the limit $a \rightarrow 0$, $N_t \rightarrow \infty$.

For the 4-particle, 3-dimensional system we are interested in, the path integral is a $3 \cdot 4 \cdot N_t$-dimensional integral, evaluated by way of the standard Metropolis algorithm. While the calculation of most quantities is straightforward, it can be seen in Eq. (\ref{eq_act}) that the kinetic energy term diverges in the limit $a \rightarrow 0$. In order to avoid this when calculating the average kinetic energy, it is more useful to use the virial estimator \cite{Herman:1982}
\begin{equation}
\braket{KE} = \frac{3}{2}NT - \frac{1}{2} \braket{\sum_{i=1}^N (\vec{x}_{ij} - \vec{\bar{x}}_i) \cdot \vec{F}_{ij}}.
\end{equation}
Here $x_{ij}$ is the position of particle i on the j-th time slice and $F_{ij}$ is the classical force on particle i at the j-th time slice. $\vec{\bar{x}}_i$ is the time-averaged center of mass of particle i. This form avoids the $a^{-1}$ divergence as $a \rightarrow 0$.

An important input of this work is the nucleon densities at kinetic freezeout for the collision energies at RHIC (see Table \ref{tab_FO}). The path integral is calculated in a volume containing one $^4 He$ nucleus or four nucleons such that $n_N V = 4$. Periodic boundary conditions are introduced via six sets of nucleons shifted in x, y, or z by a distance $L=V^{1/3}$. This gives us a total of seven boxes with the main simulation box in the center. When the center of mass of a nucleon travels outside the sides of the box, it and all of its images are shifted by $L$ to the other side of the box. See Appendix C for a discussion of the periodic box and images. 

The nuclear matter formed near freezeout in heavy-ion collisions is comprised not just of nucleons, but also large numbers of pions and other hadrons. These particles serve as the thermal bath in which the nucleons are submerged. The PIMC technique and periodic boundary conditions on the Matsubara circle incorporate effects of thermal fluctuations in a thermal bath, which e.g. in the Langevin approach are ascribed to a random force. As far a as
system in thermal equilibrium, there is no need to introduce or model collisions with
particles in the heat bath. 

The main inter-nucleon interaction that we use is (which is admittedly very simple) 
 the isoscalar central  Serot-Walecka potential \cite{Serot:1984ey}
\begin{equation}
V_{SW}(r)= -\alpha_{\sigma} \frac{e^{-m_{\sigma}r}}{r}+\alpha_{\omega} \frac{e^{-m_{\omega}r}}{r},
\end{equation}
with $\alpha_{\sigma}=6.04$, $m_{\sigma}=500$ MeV, $\alpha_{\omega}=15.17$, and $m_{\omega}=782$ MeV. While this potential describes properties of infinite nuclear matter well, it does not possess a bound state in the $4N$ system. We therefore modify slightly the repulsive term, using $\alpha_{\omega} = 11.02$ which approximately restores the desired binding energy. In Section IV, we consider the effects of modifying the potential with a reduced value of $m_{\sigma}$.  Note that this simplified potential is isoscalar, so that it is the same for $nn,np,pp$ pairs. We tested that at near-zero temperature our procedure reproduces
bound states of two and four nucleons, see 
 Appendix C.
Additionally we consider the repulsive Coulomb interaction between the two protons $V_C(r)= 7.31 \cdot 10^{-3}$ $/r$.

The PIMC simulations performed for the kinetic freezeout conditions use the temperatures listed in Table \ref{tab_FO}. For the 2.4 GeV collisions, we see from Ref. \cite{hades_freeze} that $T_{ch} = 65 \pm 1$ MeV and $T_{kin}= 71 \pm 8$ MeV. These two temperature are equal within uncertainties. Because of the smaller uncertainty and the fact that generically $T_{kin} \le T_{ch}$, we use the temperature and nucleon density calculated from the condition of chemical freezeout for this energy. All of these runs have Euclidean time discretized into $N_t= 40$ time slices. 
Simulations were performed for the six collision energies in Table \ref{tab_FO} from $\sqrt{s} = 2.4$ - $39$ GeV. For the highest energies, the low densities, and thus large box sizes, one needs to perform longer runs in order to achieve appropriate statistical accuracy in the low-$\rho$ clustering region.

\section{Physical applications}
\subsection{Preclustering in the density matrix at kinetic freezeout}
\begin{figure}[h]
	\begin{centering}
	\includegraphics[width=0.45\linewidth]{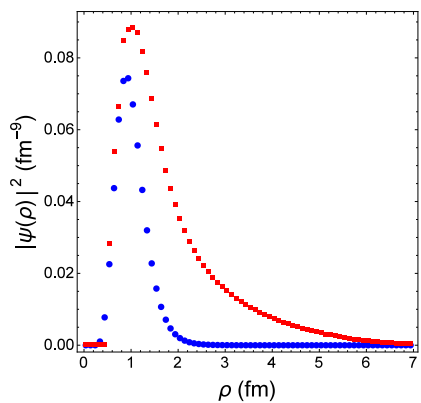}
	\caption{ Probability distributions $|\psi(\rho)|^2$ of the ground state (blue $\bullet$) and kinetic freezeout distribution (red $\blacksquare$) in hyperspace $\rho$ for $\sqrt{s_{NN}}=19.6$ GeV. }
	\label{figdist1}
	\end{centering}
\end{figure}

Calculated ensemble of paths of the four nucleons in 12 dimensions represents the
quantum/thermal density matrix of the system. In order to 
demonstrate the effect of clustering, in Fig.\ref{figdist1} 
 we show a distribution in hypercoordinate $\rho$ (\ref{eqn_rho}). The blue round points
 show the distribution at $T\approx 0$ corresponding to the ground state of the four nucleons in the alpha particle. The red squares show the distribution
at a kinetic freezeout.  While at small $\rho$ they are similar (explained by repulsive part of the potential), at large $\rho$ there appears a large tail associated with excited bound states, resonances and scattering states in the system at the corresponding temperature. 

The existence of this tail is what we call ``preclustering". More precisely, 
we associate it with the density matrix measured, minus contributions of the ground state
and the long-distance constant corresponding to propagating positive-energy states.
The probability and size of the ``precluster" 
 are the quantities we study in this work.
 
Let us discuss step-by-step how the distribution is decomposed into the three essential parts for $\sqrt{s_{NN}}=19.6$ GeV and then present the results for all energies. 

Because the ground state and kinetic freezeout distributions come from independent simulations, their relative normalization is arbitrary and they must be normalized appropriately. This is done by normalizing the kinetic freezeout distribution so that the ground state distribution fits underneath it. Fig. \ref{figdist1} shows the results of this.

Now the ground state distribution can be easily subtracted out. Next comes the less-trivial subtraction of the thermal tail. One should notice that the kinetic freezeout distribution does not become flat at long distances, as one should expect for such a distribution. This is an effect of the finite size of the box. In infinite space, one would expect $|\chi(\rho)|^2$ to go as $\rho^8$ at large hyperdistance (given that this is a 9-dimensional system with 8 hyperangular coordinates), meaning that $|\psi(\rho)|^2$ should be constant. One finds that this distribution does rise almost as $\rho^8$ for intermediate values of $\rho$, but as $\rho$ becomes comparable to the length of the box $L$, the phase space of possible configurations decreases and eventually goes to 0 at the maximum value of $\rho$ that fits in the box.  

Because of geometry of the box, 
 the thermal tail is not just a constant at large $\rho$, but rather a $\rho$-dependent distribution to be subtracted out. This distribution can be found by simply generating random configurations of four nucleons in a box of the appropriate size. 

By dividing the kinetic freezeout distribution by the random distribution, we effectively eliminate the modification of the phase space due to the geometry of the system. This reveals a distribution much more similar to the infinite-space distribution (see e.g. Fig. 5 in Ref. \cite{Shuryak:2019ikv}). 

Similarly to the ground state, the random distribution must be appropriately normalized relative to the kinetic freezeout distribution. Unlike the ground state distribution, the random distribution should not fit entirely under the kinetic freezeout distribution. The question is, to which region of the distribution should the two be equal. The answer is that it should be fit to the region of the distribution where the inter-nucleon potential $\langle V_{NN} \rangle \simeq 0$ and thus the distributions are equal. When one distribution is divided by the other, this region 
become flat as can be expected from short-range nature of the potential.
 With this normalization performed, the thermal tail, along with the ground state can be subtracted out, revealing the precluster. 

While the preclusters appear small compared to the ground states, the larger tail at large $\rho$ means that they make up a significantly larger contribution to the density matrix when the $\rho^8$ factor is included in the integration. The ground states typically make up just a few percent of the cluster, which is to be expected given the $\sim 50$ excited states that make up the precluster. Similarly, the cluster makes up a small fraction of the overall thermal distribution. Despite significant clustering being possible at these temperatures, the 4N system is still fragile, being dominated by configurations where at least one nucleon is well separated from the rest.

\begin{figure}[h]
	\includegraphics[width=0.95\linewidth]{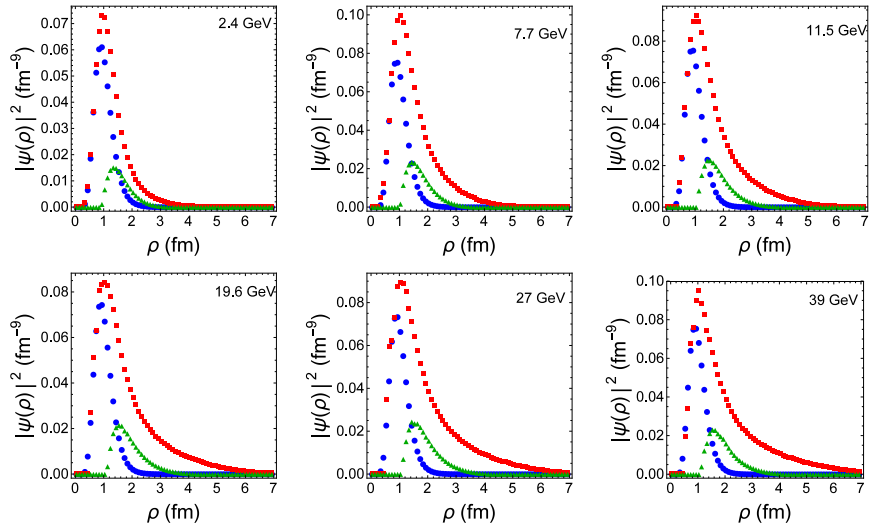}
	\caption{ Normalized probability distribution in hyperspace $\rho$ of the ground state (blue $\bullet$), kinetic freezeout distribution (red $\blacksquare$) and the precluster (green $\blacktriangle$) for the six beam energies considered}
	\label{figbig}
\end{figure}

\subsection{Angular distributions} 
One may also gain information about the density matrix by looking at the distribution in angles formed by three nucleons
\be cos(\alpha)\equiv { (\vec r_i - \vec r_j)\cdot  (\vec r_i - \vec r_k)\over |\vec r_i - \vec r_j | |\vec r_i - \vec r_k |} \ee
where $i\neq j \neq k$ are three different nucleons. Note that classical minimum of the potential energy
corresponds to  tetrahedral shape of 4-nucleon cluster, in which all sides are equilateral triangles and therefore all angles are such that 
$ cos(\alpha)=1/2$. This tetrahedral minimum-energy configuration is modified only slightly by the relatively-weak Coulomb interaction.  

\begin{figure}[h]
	\begin{centering}
\includegraphics[width=0.45\linewidth]{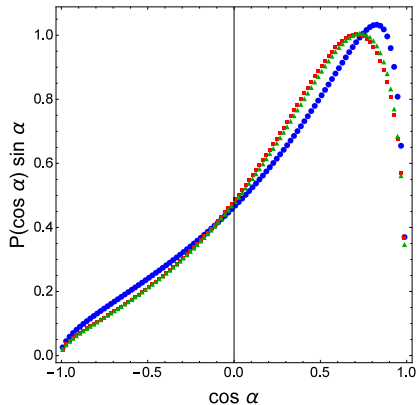}
\caption{ Normalized probability distribution of internal angles $P(\cos \alpha)$ for the ground state (blue $\bullet$), kinetic freezeout distribution (red $\blacksquare$), and the random distribution (green $\blacktriangle$) for conditions at $\sqrt{s_{NN}} = 19.6$ GeV.}
\label{figangles}
	\end{centering}
\end{figure}

In Fig. \ref{figangles} we see, as before, that the angular distribution of the ground state is both broad and $not$ centered at $cos(\alpha) = 1/2$. The broad distribution is the result of quantum/thermal fluctuations around the minimum-energy configuration stemming from the kinetic term in the action. The shift in the peak to an angle near $36^\circ$ is due partially to the Coulomb interaction of the protons and also due to the fact there is simply more phase space to produce a triangle with small angles than large ones, as can be seen by the random distribution.

The kinetic freezeout distribution is nearly identical to that of randomly-placed nucleons. This makes sense as the system spends most of its time at large values of $\rho$ outside of the clustering region where the average inter-nucleon potential $\langle V_{NN} \rangle \simeq 0$. The cluster at kinetic freezeout, which is a sum of states of various angular momenta and the thermal tail, shows no angular correlation between the nucleons. The distributions in Fig. \ref{figangles} show no beam-energy dependence. The wide distribution in internal angles, even in the ground state, shows the importance of an approximation-free method for few-body quantum systems such as PIMC. 
 
\subsection{Modification of the inter-nucleon potential} 
 In heavy-ion collisions the medium is expected to modify the parameters of the inter-nucleon potential. This is especially true near the critical point, where long-range correlations should arise. The most dramatic effect suggested of this form is the reduction of the $\sigma$ mass $m_{\sigma}$ near $T_c$ \cite{Stephanov:1998dy}
 \begin{equation}
 m_{\sigma} \sim \left( \frac{|T-T_c|}{T_c}\right)^{\nu}.
 \end{equation}
 Such a reduction in the $\sigma$ mass, which drives the attractive portion of the interaction, will greatly increase the strength and range of the attraction. 
\begin{figure}[h]
	\begin{centering}
	\includegraphics[width=0.45\linewidth]{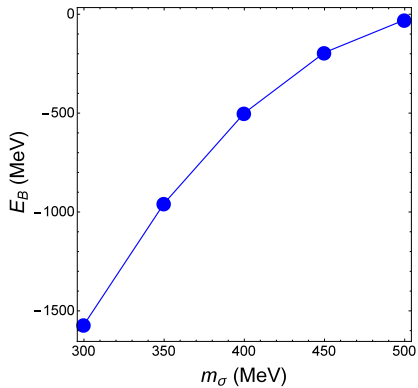}
	\caption{Binding energy $E_B$ of the 4N system as a function of the $\sigma$ mass $m_{\sigma}$}
	\label{figbind}
	\end{centering}
\end{figure}

Clearly such a modification will greatly modify the clustering dynamics of the 4N system of interest in this work. The most obvious question to ask is then: how much modification of $m_{\sigma}$ should be expected near kinetic freezeout? The answer is that it should be only a small modification for two main reasons. The first is that the reduction of $m_{\sigma}$ to 0 is expected at $T_c$ which is some $\sim 40$ MeV above $T_{kin}$ for a wide range of beam energies. The second is that even at the critical point, the finite size and finite lifetime of the QGP system prevent the correlation length from getting too large. 

In this section, we consider a modification of the system in which the binary interactions are modified with a reduced $m_{\sigma}$. We should note that this is certainly not a realistic description of the inter-nuclear forces that occur near CP. As shown in Ref. \cite{DeMartini:2020anq}, fluctuations near CP generate repulsive 3- and 4-body forces which grow as large powers of the correlation length $\xi$, and for large enough $\xi$ may \textit{suppress} cluster production. The results in this section should be seen as describing how a modification of the potential can affect clustering, but not as a realistic model of the forces near CP. 

The strengthening of the attractive force is seen clearly in Fig. \ref{figbind}. Here the binding energy of the 4N system is shown for the $\sigma$ mass reduced down to $300$ MeV. Reducing the mass down to this value increases the binding energy from its physical value of $28.3$ MeV by approximately a factor of 50. For the smallest value of $m_{\sigma}$ studied, 300 MeV, the binding energy is greater than 1.5 GeV. Such a deeply-bound state is unrealistic. This is a limit of the simplicity of our model. 

\begin{figure}[h]
	\begin{centering}
	\includegraphics[width=0.45\linewidth]{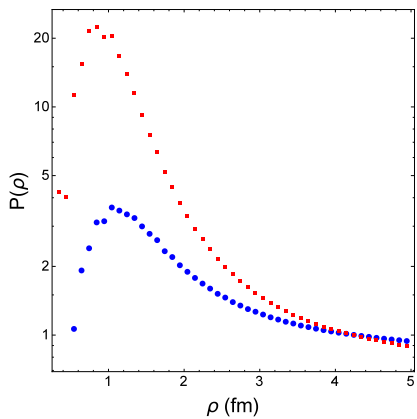}
	\caption{ Probability distribution relative to random distribution $P(\rho)$ at the kinetic freezeout conditions for $\sqrt{s} = 19.6$ GeV with $m_{\sigma} = 500$ MeV (blue $\bullet$) and $m_{\sigma} = 450$ MeV (red $\blacksquare$)}
	\label{figmod}
	\end{centering}
\end{figure}

The effect of the modified interaction on clustering is seen directly in Fig. \ref{figmod}. Reducing the standard $\sigma$ mass by 50 MeV causes the peak correlation to jump by a factor of ~5. Clearly, even modest modifications of the forces can strongly affect clustering.

\section{Estimates of the observables} 
\begin{table}[t]
\begin{center}
\caption{Low-lying resonances of the $^4$He system, from BNL properties of nuclides 
$J^P$ are the total angular momentum and parity,
$\Gamma$ is the decay width. The last column is the decay channel branching ratios, in percent. 
$p,n,d$ correspond to the emission of proton, neutron, or deuterons respectively.
}
\begin{tabular}{|c|c|c|c|}
\hline
$E$ (MeV) & $J^P$ & $\Gamma$ (MeV)  & decay modes, in \% \\
\hline
\hline
20.21 & $0^+$ & 0.50  & $p$ = 100\\
21.01 & $0^-$ & 0.84  & $n$ = 24,  $p$ = 76\\
21.84 & $2^-$ & 2.01  & $n$ = 37,  $p$ = 63  \\
23.33 & $2^-$ & 5.01  & $n$ = 47,  $p$ = 53  \\
23.64 & $1^-$ & 6.20  & $n$ = 45,  $p$ = 55 \\
24.25 & $1^-$ & 6.10  & $n$ = 47,  $p$ = 50,  $d$ = 3 \\
25.28 & $0^-$ & 7.97  & $n$ = 48,  $p$ = 52\\
25.95 & $1^-$ & 12.66 & $n$ = 48,  $p$ = 52 \\
27.42 & $2^+$ & 8.69  & $n$ = 3,   $p$ = 3,   $d$ = 94 \\
28.31 & $1^+$ & 9.89  & $n$ = 47,  $p$ = 48,  $d$ = 5 \\
28.37 & $1^-$ & 3.92  & $n$ = 2,   $p$ = 2,   $d$ = 96 \\
28.39 & $2^-$ & 8.75  & $n$ = 0.2, $p$ = 0.2, $d$ = 99.6 \\
28.64 & $0^-$ & 4.89  & $d$ = 100 \\
28.67 & $2^+$ & 3.78  & $d$ = 100 \\
29.89 & $2^+$ & 9.72  & $n$ = 0.4, $p$ = 0.4, $d$ = 99.2 \\
\hline
\end{tabular}
\label{tab:he4}
\end{center}
\end{table}

\subsection{Decay channels of the 4N system}

The ``preclusters" identified by our calculation are statistical correlations in the density matrix
at the kinetic freezeout. Since, by definition,  no collisions happens after that, the four nucleons
are either kept together, in the ground and excited bound
 states, or get separated into various decay channels. 

The set of $known$ excited four-nucleon states is reproduced in the Table I.
Note that the energy here is the excitation energy, counted from the ground state.
Since binding is -28.3 MeV, the highest state (at the bottom of the table) has absolute energy 
of about 1 MeV above zero. Note however, that its width is an order of magnitude larger
than that, and also larger than the $\sim$ 1 MeV spacing between levels. Clearly the 
table is terminated because of experimental difficulties to identify overlapping resonances, rather than
by  the actual existence of these resonances. 

The main experimental observables of the clustering of the 4N system is the yield of bound $^4He$, and yields of the
 decay products of other bound states and resonances. One of the goals of this work is to determine dynamically rather than statistically the fraction of these correlated clusters which decay into the ground state. Without knowledge of the wave functions of the various excited states, the precluster is assumed to populate the various states in Table \ref{tab:he4} statistically. These excited states then decay into light nuclei with probabilities that are known experimentally. Note that only one decay product is known for each decay mode. We assume all decays listed in Table \ref{tab:he4} are two-body decays. In principle, the system could decay into three nuclei ($p$+$n$+$d$) or four nucleons ($p$+$p$+$n$+$n$). 

\begin{table}[h]
	\caption{Decay probability (in percent) of the 4N cluster into $^4He$ or two-body states for the different collision energies. Three- and four-body decays are not considered here.}
\begin{center}
\begin{tabular}{|c|c|c|c|c|}
\hline
$\sqrt{s}$ (GeV) & $^4He$ & $p$ + $t$ & $n$ + $^3He$ & $d$ + $d$ \\
\hline
\hline 
2.4 & 15.8 & 25.8 & 19.0 & 39.4 \\
7.7 & 7.3 & 28.0 & 20.7 & 44.0 \\
11.5 & 3.7 & 29.1 & 21.5 & 45.8 \\
19.6 & 2.8 & 29.3 & 21.7 & 46.2 \\
27. & 2.5 & 29.4 & 21.8 & 46.3 \\
39. & 2.4 & 29.5 & 21.8 & 46.3 \\
\hline 

\end{tabular}
\end{center}
\label{tabDecay}
\end{table}

At all but the lowest energy studied, the ground state makes up only a few percent of the total cluster. Only at $\sqrt{s}=2.4$ GeV is the probability to decay to $^4He$ comparable to the individual two-body decay probabilities. This is due to the significant reduction in $T_{kin}$ at this energy increasing the relative statistical weight of the ground state. Finite-density effects are much smaller than the effect of changing temperature. For the five energies with roughly equal $T_{kin}$ increasing the density modestly increases the ground state probability. This may be attributable to the size of the box being comparable to or smaller than the size of some of the excited states, removing them from the spectrum of allowed states, while the smaller ground state does not feel the effects of the boundary. 

Decays of the 4N system should contribute only a small amount to the total yield of these particles. A complete picture of light nuclei production should include statistical hadronization as well as feed-down from excited nucleon states (such as $N^*$ or $\Delta$), which should greatly outnumber 4N states.  

\subsection{Virial expansion and clustering}
The potential part of the partition function (of single species system) of $N$ particles  can be re-written in the form 
\begin{equation} Z_{pot}=1+{1 \over V^N} \int d^3 x_1...\int d^3 x_N \big[ e^{\big(-\sum_{i>j} V(\vec x_i -\vec x_j)/T\big) } -1\ \big] 
\end{equation}
by adding and subtracting 1. Since we focus on clusters of 4 particles, coordinates of all others can be integrated out, as well as the coordinate of
its center of mass. One may study the $2N$ and $3N$ systems to determine the size of the effect proportional to $n$ and $n^2$, respectively. However, given the small binding of $d$ and $t$, one should expect such contributions to be small.  What is left is
\begin{eqnarray}  Z_{pot}&=&1+{N(N-1)(N-2)(N-3) \over  4! }({V^{(9)}_{cor} \over V^3}) \nonumber \\
&\approx& 1+ n^3 V^{(9)}_{cor}{N \over 4!} 
\label{eqVir}
\end{eqnarray}
where the so-called 9-dimensional  correlation volume $V^{(9)}_{cor}$ is
\begin{equation}
V_{cor}^{(9)} = \frac{32}{105}\pi^4 \int d\rho \rho^8 (P(\rho)-1),
\end{equation}
Here $P(\rho)$ is the probability distribution relative to an ideal gas in 9-dimensional radius $\rho$  
and the factor in front is the solid angle in 9 dimensions. We neglect repulsion and integrate over the region in which the integrand is positive. 
The addition to free energy is then
$ \Delta (-T log(Z))= - T n^3 V^{(9)}_{cor}{N \over 4!} $ , same as to the grand partition sum. Differentiating it over $\mu$, present in each $N$,
one finds the addition to the particle number $ \Delta N/N=  n^3 V^{(9)}_{cor}/3! $ (a factor of 4 cancels out). 

As a check on the numerical factor in the thermodynamic expression, we can compare it to a more 'direct' method of computing the ratio of clusters. We can define the total volume of the entire distribution $V^{(9)}_{tot}$ analogously and then compute to ratio $R$ of clusters to unclustered configuration $R= V^{(9)}_{cor}/(V^{(9)}_{tot}-V^{(9)}_{cor})$. 

The expressions are  modified in the case of several particle species. In the problem at hand, nucleons have spin 1/2 and isospin 1/2,
so the number of distinct species is 4. If $n_s$ is density {\em per one species}, the total density of symmetric matter is simply $n_B=4n_s$. In the case of particular clusters we actually simulate, made
of four distinct species,   there is no need for symmetrization and there is no $4!=24$ in the denominator. However,  the density in front is in this case $n_s$, not total $ n_B$, 
and the numerical suppression factor is actually $1/4^3=1/64$. Therefore, the
clustering contribution to $\langle N \rangle$ is small, at the percent level or less.

\begin{table}[h]
	\caption{Correlation volume $V_{cor}^{(9)}$ of the 4N system at all beam energies for $m_{\sigma} = 500, 450, 400$ MeV.}
	\begin{center}
		\begin{tabular}{|c|c|c|c|}
			\hline
			$\sqrt{s}$ (GeV) & \multicolumn{3}{c|}{$V_{cor}^{(9)}$ (fm$^9$)} \\	
			\hline
			\hline
			$m_{\sigma}$ (MeV) & $500$ & $450$ & $400$ \\
			\hline
			2.4 & $8.7 \cdot 10^4$  & $4.4 \cdot 10^5$ & $4.5 \cdot 10^6$ \\
			7.7 & $4.3 \cdot 10^4$ & $1.9 \cdot 10^5$ & $1.6 \cdot 10^6$ \\
			11.5 & $8.8 \cdot 10^4$ & $2.2 \cdot 10^5$ & $9.0 \cdot 10^5$ \\
			19.6 & $7.2 \cdot 10^4$ & $2.7 \cdot 10^5$ & $7.6 \cdot 10^5$ \\
			27. & $9.5 \cdot 10^4$ & $2.5 \cdot 10^5$ & $7.1 \cdot 10^5$ \\
			39. & $9.1 \cdot 10^4$ & $2.4 \cdot 10^5$ & $6.9 \cdot 10^5$ \\
			\hline
			
		\end{tabular}
	\end{center}
	\label{tabvol}
\end{table}

\begin{figure}[h]
	\begin{center}
		\includegraphics[width=0.45\linewidth]{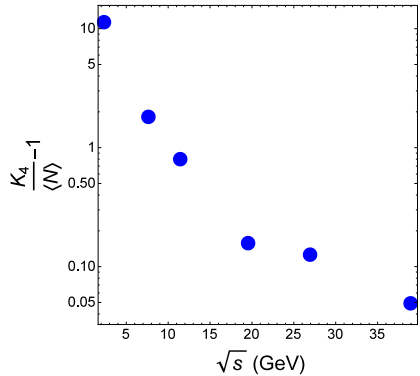}
		\caption{The 4th cumulant deviation (Eq. (\ref{eqn_K4toK1m}))
			versus $\sqrt{s}$, using the 9-dimensional correlated volume $V^{(9)}_{cor}$ determined from the PIMC simulations.}
		\label{fig_K4toK1m}
	\end{center}
\end{figure}

Furthermore, the $n$-th derivative of $log(Z)$ over chemical potential, called $K_n$, can also be calculated.
One finds that $K_4$, with extra three derivatives
compared to $K_1=\langle N \rangle$, does $not$ have this $1/4^3$ numerical factor, and so, in the same approximation as above,
\begin{equation} \label{eqn_K4toK1m}
{K_4 \over \langle N \rangle}-1= n_B^3 V^{(9)}_{cor}
\end{equation}
The values of the r.h.s. are shown in Fig. \ref{fig_K4toK1m}. As one can see,
the predicted
cluster contribution to the 4th moment are no longer small, for the two left points,
corresponding to HADES and the lowest BES-I energy $\sqrt{s}=7.7\, GeV$.
While the experimental values of the 4-th moment of proton distribution at these energies still
have relatively large errors, the observed deviation from zero is substantial. 

Another perspective at the observed cumulants $K_n$ can be obtained using  $factorial$ cumulants $C_n$. In particular
\begin{equation}
K_4 - \langle N \rangle= 7C_2+6C_3+C_4.
\end{equation}
According to Ref.
\cite{Koch:2020azm}, BES-I data show small values for $ C_2,C_3$ and therefore the moment $K_4$
is in fact dominated by the nonzero factorial cumulant $C_4$. 
It correlates well with our general finding, that, 
under freezeout conditions, the nucleon clustering starts from the 4N systems.
We conclude emphasizing that measurements of $K_4$, $C_4$ is related to clustering, 
and therefore  their  dependence of  on the collision energy is of great interest.

Characterizing the strength of spatial correlations in the 4N system is necessary for making predictions of the overall magnitude of the feed-down contributions to light nuclei
yields. Models of light nuclei production, such as the previously-mentioned coalescence model \cite{Sun:2018jhg} show explicit dependence on spatial correlations of nuclei. Our correlated value is a measure of such correlations. 

From Tables \ref{tabDecay} and \ref{tabvol2} one can then estimate that about $25\%-30\%$
of the clusters decay into $t$. The resulting ratio of tritium to proton yields $t/p$ is then about $0.4\%$ at $\sqrt{s} = 7.7$ GeV, comparable to the STAR BES data. See Ref. \cite{Zhang:2020ewj} for recent STAR data on light nuclei yields. 

Table \ref{tabvol} lists the values of $V_{cor}^{(9)}$ for all six beam energies and for three values of $m_{\sigma}$. These values should only be taken with a moderate level of uncertainty. The correlated volume is sensitive to the method of subtracting out the long-distance thermal tail from the distribution. Even a change of a few percent in the normalization factor (which comes from averaging over the flattest portion of the distribution $P(\rho)$) can lead to a factor of 2 change in $V_{cor}^{(9)}$. It is clear that the largest source of uncertainty (and room for greatest improvement) comes from this subtraction of the thermal tail. Further studies should not solely rely on hyperdistance $\rho$, but rather distributions in the full set of 9-dimensional hypercoordinates. 

\begin{table}[h]
	\caption{Ratio of clusters to unclustered configurations calculated from the thermodynamic contribution expected in Eq. (\ref{eqVir}) and the ratio $R$ computed from direct integration over distributions.}
	\begin{center}
		\begin{tabular}{|c|c|c|}
			\hline
			$\sqrt{s}$ (GeV) & $\frac{n_B^3}{64} V^{(9)}_{cor}$ & $R$ \\
			\hline
			\hline
			2.4 & .180  & .181 \\
			7.7 & .0288 & .0371 \\
			11.5 & .0127 & .00193 \\
			19.6 & .0025 & .0073 \\
			27. & .0020 & .0045 \\
			39. & .00078 & .0021 \\
			\hline
			
		\end{tabular}
	\end{center}
	\label{tabvol2}
\end{table}

The values of $V_{cor}^{(9)}$ increase significantly as the attractive inter-nucleon force is strengthened, even modestly. Thus, observables such as the light nuclei ratios mentioned, which should be affected by clustering, can serve as signals of potential modifications of inter-nucleon forces. 

\section{Nucleon Multiplicity Distribution}

While the total baryon number is conserved in collisions, the observed multiplicity distribution shows fluctuations which may be caused by many effects: (i) wandering inside and outside the detector acceptance; (ii) turning of the observable protons into unobservable neutrons; (iii) turning into light nuclei species, and, last but not least, (iv) the preclustering phenomenon. The first two are well studied with various event generators. Anti-correlations between proton and light nuclei multiplicities in individual bins are still to be experimentally studied. 

Here we give two reasons why the preclustering phenomenon affects the nucleon multiplicity distribution. The first, the so-called 'volume' effect, can be explained as follows. Suppose a certain number of preclusters of four or more nucleons are formed at freezeout. Instead of focusing on their two-body decays to light nuclei as we did before, let us think about feed down into four individual nucleons.

On one hand, the number of preclusters is relatively small, and their branching ratio to such modes is also not large, so one may expect that such feed down is completely negligible. This is indeed the case near the maximum of the multiplicity distribution $N_p \simeq \langle N_p \rangle$. Yet the effect should be large and observable \textit{at the tails} of the multiplicity distribution. The main multiplicity distribution is, to zeroth approximation, just a Poisson distribution, with the mean defined by detector acceptance. Its tails, at small and large multiplicity, are the probability that a large number of protons happen to cross the acceptance boundary and be either outside or inside it. For example, according to STAR data \cite{Luo:2017faz,Bzdak:2018uhv} for central $Au-Au$ collisions at $\sqrt{s} = 7.7$ GeV we will refer to, $\langle N_p \rangle \simeq 40$ and the tails we discuss are at $N_p \sim 10$ and $N_p \sim 70$. Another way to say it is that the distribution in cluster number, also approximately given by a Poisson distribution, is wider because \textit{the mean number of clusters is smaller}. While the average number of clusters is small, the wide distribution means they can contribute significantly at the tails, particularly the low $N_p$ tail.  

Indeed such deviations from Poisson are observed. In order to quantify those, one now uses a sensitive statistical tool, calculating the \textit{factorial cumulants} of the distributions. Recall that for the Poisson distribution $C_i = 0$ for $i > 1$. For the STAR data in question, they are instead (the maximum values taken from all energies and centralities)
$$C_2\approx -2, \,\,\, C_3\approx -10,\,\,\,  C_4\approx 175.$$

While the exact reason for these large values of $C_3$ and $C_4$ remain unknown, we present here an \textit{ad-hoc} model, which quantifies the effect and qualitatively reproduces the cumulants. (Its discussion was inspired by another model proposed in Ref. \cite{Koch:2020azm}.) It has the distribution 

\begin{equation}
P(N_p) = 0.995P_P(N_p,40)+0.005P_P(N_c,9.1)
\label{eqpoisson}
\end{equation}
where $P_P(N,\langle N_p \rangle)$ is the Poisson distribution and $N_c$ is the number of clusters of four protons. This particular example gives $C_3 = -5.2$ and $C_4 = 190$.

\begin{figure}[h]
	\begin{centering}
	\includegraphics[width=0.45\linewidth]{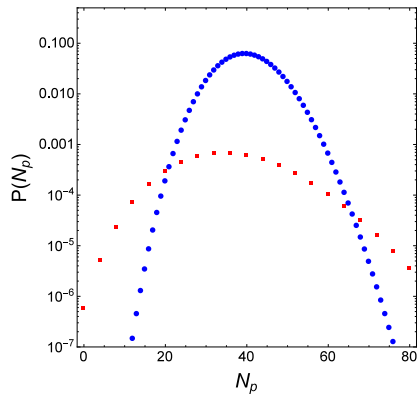}
	\caption{ The two Poisson components of Eq. (\ref{eqpoisson}), proton component (blue $\bullet$) and $4N$ cluster component (red $\blacksquare$)}
	\label{figpoisson}
	\end{centering}
\end{figure}

Of course, this is just a schematic model. An assumption that a cluster decays into exactly 4 protons is unrealistic as some of them are neutrons. That is why we took the cluster probability normalization to be so low, 0.005.

\begin{figure}[h]
	\begin{centering}
	\includegraphics[width=0.45\linewidth]{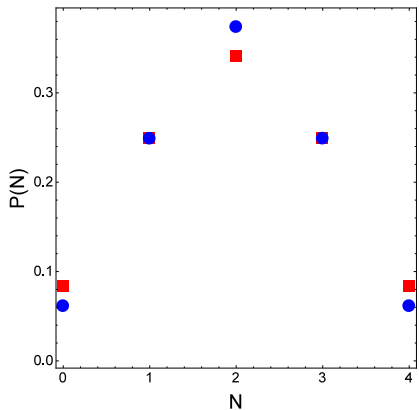}
	\caption{ Probability distribution of number of nucleons N in half of simulation box for a random distribution (blue $\bullet$) and for interacting nucleons at kinetic freezeout conditions (red $\blacksquare$) at $\sqrt{s} = 2.4$ GeV with unmodified $\sigma$ mass.}
	\label{figmulti}
	\end{centering}
\end{figure}

Another effect, which we call the 'boundary' effect, so far ignored is that clusters are treated like point-like objects, so they are either in or out of the observable phase space for all nucleons. The clusters do have certain coordinate and momentum spreads in the density matrix, so they may be outside of the volume \textit{partially}. This effect can be quantified using ensembles from our simulations. Dividing the volume in half along any axis, we record the probability for observing $N$ nucleons ($0-4$) in said half. The result is shown in Fig. \ref{figmulti}, showing that the probability distribution for interacting nucleons is wider than randomly-placed nucleons. 
  
This, of course, modifies the nucleon multiplicity distribution, specifically by increasing the probability that either all or none of the nucleons in the 4N system are in the phase space accessible to the detector. This effect could be greatly enhanced or suppressed for modified interactions stemming from fluctuations near CP. In fact, it is precisely such modification of the NN interaction which has generated such interest in proton multiplicity distributions.

The definitive feature of the QCD critical point is the large increase in correlation length $\xi$ of the QGP system. Given that the correlation length may only increase modestly at freezeout in heavy-ion collisions, much work has been done on identifying observables which depend more strongly on $\xi$. It has been proposed \cite{Stephanov:2008qz,Asakawa:2009aj} that non-Gaussian moments of the proton multiplicity distribution display such behavior. In particular, the kurtosis $\kappa_4$ is expected to be sensitive to $\xi^7$. Much experimental effort has been exerted to study these distributions including at the programs discussed in this work, HADES and STAR \cite{Adamczyk:2013dal,Adamczewski-Musch:2020slf}. 

\section{Summary}

A program of quantum/statistical mechanical studies of few-nucleon clustering has been started
by Refs. \cite{Shuryak:2019ikv,Shuryak:2018lgd}. The theoretical methods used in those papers include classical molecular dynamics,
semiclassical ''flucton" approach at finite temperatures, and  hypercoordinates in $3(N-1)$-dimensional Jacobi coordinate space.
They reveal a significant amount of ``preclustering" at the kinetic freezeout stage of heavy ion collisions and put emphasis on $four-nucleon$
systems, as the lightest one possessing multiple states/resonances. 

The goal of this paper is to check calculations using this set of approximate methods by a direct first-principle calculation based on path-integral Monte Carlo.
Our main results are shown in Fig. \ref{figbig}, in which we plotted the density matrix in terms of the hyper-radial coordinate $\rho$.  
To the extent we can compare those results with those of approximate methods mentioned, we see a rather consistent picture.

One can see that the precluster shape is different from that of the ground wave function squared, so that after freezeout quantum mechanical
decomposition of the preclusters should produce not only $^4He$, but also a superposition of (near-zero-energy) bound and resonance states.
Those have close energies but different quantum numbers (in particular, angular momenta), and therefore have different decay widths  and branching
ratios. Fortunately, these states and their decays were experimentally studied long ago, so in principle one can evaluate the {\em feed-down} from them
into yield of light nuclei, such as $d$, $t$, $^3He$, and $^3_{\Lambda}He$. Recent summaries of experimental situation can be found in \cite{Oliinychenko:2020ply,   
Shuryak:2020yrs}. The most intriguing experimental observation is that the yield ratio $N_t N_p/N_d^2$ seems to non-monotonously depend on
collision energy, with apparent maximum at BES-1 mid-range $\sqrt{s}\sim 20\, GeV$. The value of the ratio, especially from low energy HADES data,
are very different from ratio of statistical weights, indicating presence of strong feed-down into tritium. 

Another manifestation of clustering can potentially be studies of moments of proton multiplicity distribution. Using the 9-dimensional correlation volume evaluated from PIMC data, we calculated the fourth-order virial coefficient of nucleon matter at kinetic freezeout. While its contribution to particle number is small, at the percent level, we found that its contribution to $C_4/\langle N \rangle$ becomes of order one at collision energy $\sqrt{s}=7.7$ GeV, see Fig. \ref{fig_K4toK1m}, as was indeed observed by STAR. We also suggest that clustering enhances the low-$N$ tail of the multiplicity distribution and propose a model for qualitatively reproducing STAR data.

If there exists a QCD critical point, one expects  additional contributions to those from fluctuations of the critical mode \cite{Stephanov:1998dy}
when the freezeout occurs close to its location on the phase diagram.  Nonlinear features  of such fluctuations are complicated,
but fortunately well known e.g. from lattice studies of the Ising model. The preclustering phenomenon will be modified by these fluctuations. 

\section*{Acknowledgments}
	This work is supported by the Office of Science, U.S. Department of Energy under Contract No. DE-FG-88ER40388. We also thank the Stony Brook Institute for Advanced Computational Science for providing computer time on its cluster. 

\appendix

\section{Bulk properties of matter at the freezeouts}
The RHIC beam energy scan, suggested to look for QCD critical point and whose first results were reported in Ref. \cite{Mohanty:2011nm}, took data at the energies listed in Table \ref{tab_FO}. 

\begin{table}[t]
\caption{The collision energies of RHIC in low energy scan; 
Fitted chemical freezeout parameters, $T_{ch}$,  $\mu^B_{ch}$, and   $R_{ch}$,  from Grand Canonical Ensemble fit to particle yields \cite{Adamczyk:2017iwn},
for the most central bins $[0,5\%]$; The temperatures of kinetic freezeout    $T_{kin}$ 
from blast wave fit to STAR spectra  \cite{Adamczyk:2017iwn}; Parameters for $2.4$ GeV collisions are taken from analogous analyses of HADES data \cite{hades_freeze}; Nucleon corrected chemical potential at freezeout by Eq. (\ref{eqn_mu_final}) and corresponding thermal nucleon densities $n^B_{kin}$.}
\begin{center}
\begin{tabular}{|c|c|c|c|c|c|c|}
\hline
$\sqrt{s}$ (GeV)       & 2.4  & 7.7     &  11.5   &  19.6 & 27.      & 39. \\
\hline
\hline
 $T_{ch}$ (MeV)       & 65.  & 143.8 & 150.6 & 157.5 & 159.8  & 159.9 \\
  $\mu^B_{ch}$ (MeV) & 784 &  398.2 & 292.5 & 195.6  & 151.9 & 104.7  \\
  $R_{ch}$ (fm)          & 9.6  & 5.89    & 6.16   &  6.04  & 6.05   & 6.27  \\
   $T_{kin}$ (MeV)       & 71. & 116.  &  118.  &   113.  & 117.   &  117.  \\
     $\mu^B_{kin}$ (MeV) & - &  503. & 432. & 406. & 363. & 329. \\
$n^B_{kin}$ (fm$^{-3}$) & 0.051 & 0.035 & 0.021 & 0.013 &  0.011 & 0.0082 \\
\hline
\end{tabular}
\end{center}
\label{tab_FO}
\end{table}%

As one can see from the table, in the scan energy region the temperature 
of chemical freezeout  $T_{ch}$ is growing (errors $\pm 3\, MeV$) to a constant,
while the baryon chemical potential  $\mu^B_{ch}$ strongly decreases. 
The fireball volume within this RHIC energy scan range approximately doubles.
The  temperature of the kinetic freezeout  $T_{kin}$ defined from ``blast wave" fit, is, on the other hand,  constant within errors ($\pm 11\, MeV$). The mean 
flow velocity is also constant $\langle\beta\rangle\approx 0.46 \pm 0.04$. 

The change in thermal state of hadronic matter between chemical and kinetic freezeout we treat following Ref. \cite{Hung:1997du}.
For particles other than pions the expression for it is given by
\be \mu(T)= \mu_{ch}{ T \over T_{ch}} + m (1 - {T \over T_{ch}}) \label{eqn_mu_final} \ee 
The corresponding values for nucleons at freezeout are given in the table,
together with the thermal baryon densities at kinetic freezeout. Those are the ones used in
simulations described in the main text.

For completeness, let us mention that for pion the chemical potential at kinetic freezeout 
at $T_{kin}\approx 117\, MeV$ is $\mu_\pi \approx 62\, MeV$, by a curve given in Ref. 
\cite{Teaney:2002aj}. The modification of thermal pion spectrum at small $p_t$ induced by pion chemical potential (of similar 
 magnitude) has been demonstrated already in the original paper  \cite{Hung:1997du}, using the pion spectra from SPS NA44 experiment. This effect was recently reconfirmed in LHC  ALICE
 pion spectra,
 see Ref. \cite{Begun:2013nga}.

\section{Feed down from excited nucleon states} 

The thermal conditions of the fireball produce, in principle, all species of hadrons with some non-zero density, most of which decay long before reaching detectors. This includes excited states and resonances. Of particular interest to this work are the excited nucleons states which decay into $p$ and $n$. At chemical freezeout, inclusion of such states, as well as even weak decays, is necessary to get accurate fits of particle yield ratios \cite{Andronic:2005yp}.

After chemical freezeout the numbers of individual species of hadrons are fixed as inelastic collision cease. Due to the system expanding however, the density decreases. Decays from excited states may increase the number of $p$ and $n$, however. The time at which these resonances decay after a collision is still an open question and different models assuming different ordering of events (resonance decay \textit{then} nuclei coalescence or vice-versa) give different prediction for the light-nuclei ratio $N_tN_p/N_d^2$ \cite{Oliinychenko:2020ply}. In this work, it has been assumed that feed down from these states is not significant by the time kinetic freezeout occurs and the nucleon densities used are computed directly from $T_{kin}$ and $\mu_{kin}^B$. One should expect that if such feed down substantially increases the nucleon density at kinetic freezeout, the role of clustering would be more significant. 

At the lower temperatures of kinetic freezeout, one would expect a reduced contribution from these $ > 1$ GeV particles. We have calculated, using the textbook statistical formulas, the ratio of the densities of these states to the nucleon density at kinetic freezeout for a few of the collider energies.    
\begin{table}[h]
	\caption{Ratio of densities of nucleon excited states to the nucleon density at kinetic freezeout conditions for all $N^*$ and $\Delta$ states with mass $\le 1.7$ GeV assuming the states have not yet decayed. Number in parentheses represents mass of the state in MeV; $J^P$ is total angular momentum and parity. List of nucleon states taken from Ref. \cite{Tanabashi:2018oca}.}
\begin{center}
	\begin{tabular}{|c|c|c|c|c|}
		\hline
		$N^*$ state& $J^P$ & 2.4 GeV & 7.7 GeV & 39 GeV\\
		\hline
		\hline
		$N(1440)$ & $\frac{1}{2}^+$  & $8 \times 10^{-4}$ & $0.023$ & $0.024$ \\
		$N(1520)$ & $\frac{3}{2}^-$ & $5\times 10^{-4}$ & $0.025$ & $0.026$ \\
		$N(1535)$ & $\frac{1}{2}^-$ & $2\times 10^{-4}$  & $0.011$ & $0.012$ \\
		$N(1650)$ & $\frac{1}{2}^-$ & $4\times 10^{-5}$ & $0.005$  & $0.005$ \\
		$N(1675)$ & $\frac{5}{2}^-$ & $8\times 10^{-5}$ & $0.011$ & $0.012$ \\
		$N(1680)$ & $\frac{5}{2}^+$ & $7\times 10^{-5}$ & $0.011$ & $0.012$ \\
		$N(1700)$ & $\frac{3}{2}^-$ & $4\times 10^{-5}$ & $0.006$ & $0.007$ \\
		$\Delta(1232)$ & $\frac{3}{2}^+$ & $0.033$ & $0.23$ & $0.23$ \\
		$\Delta(1600)$ & $\frac{3}{2}^+$ & $2\times 10^{-4}$ & $0.014$ & $0.014$ \\
		$\Delta(1620)$ & $\frac{1}{2}^-$ & $6\times 10^{-5}$ & $0.006$ & $0.006$ \\
		$\Delta(1700)$ & $\frac{3}{2}^-$ & $4\times 10^{-5}$ & $0.006$ & $0.007$ \\
		\hline
		Total & - & 0.035 & 0.348 & 0.355 \\
		\hline
	\end{tabular}
\end{center}
\end{table}
 
There is a clear $\sqrt{s}$-dependence, with the excited states having a much-reduced relative density at 2.4 GeV due to the reduced temperature. In all cases, the lightest resonance considered, $\Delta(1232)$ has a higher density than all other excited states considered here combined. Throughout the range $\sqrt{s} \sim 7.7 - 39$ GeV, the total density of the excited states should be about 30\% of the nucleon density. For comparison (see Table \ref{tabvol2}), the statistically-correlated clusters studied in out simulation make up about 1\% of the total configurations sampled over the same energy range. This indicates that the total amount of feed down from the $4N$ system should be comparatively small. At $\sqrt{s} = 2.4$ GeV however, our results indicate that the density of clusters should be much higher than that of the excited nucleon states, confirming the importance of feed down from the clusters at low beam energies.  

\section{The ground states of two and four nucleons} 
The first task is to tune PIMC parameters and to test whether the ground state binding and the wave functions are correctly reproduced. The code was tested by producing the ground state probability distribution of the Walecka deuteron - two nucleons interacting via the Serot-Walecka potential. Being a two-body system with only a radial interaction, the Schrödinger equation can be solved without approximation. 

\begin{figure}[h]
	\begin{centering}
	\includegraphics[width=0.45\linewidth]{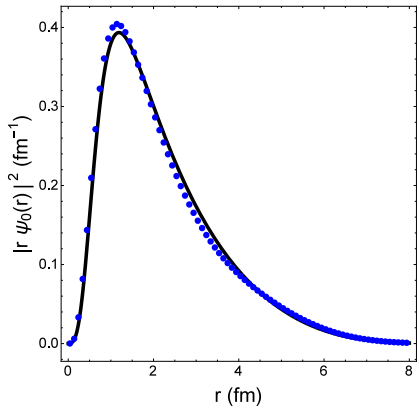}
	\caption{Normalized probability distribution for the Walecka deuteron. The curve is the numerical solution of the Schrödinger equation.}
	\end{centering}
\end{figure}

Calculations for all ground state distributions of the 4N system were done at $T= 0.5$ MeV, a temperature that is sufficiently small compared to the energy of the first excited state of $^4 He$ (see Table I). This leads to a Matsubara time of 394 fm, discretized into $N_t= 3000$ time steps. 

\begin{figure}[h]
	\begin{centering}
	\includegraphics[width=0.45\linewidth]{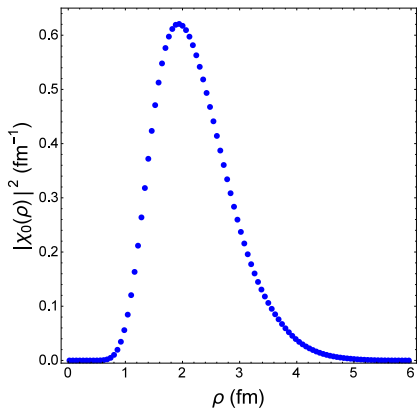}
	\caption{Normalized probability distribution of the ground state $|\chi_0(\rho)|^2$ of $^4 He$ over hyperdistance $\rho$.}
	\label{fig_ground}
	\end{centering}
\end{figure}

The system of four interacting nucleons exists in a 9-dimensional space. In order to study the dynamics of the system, it is useful to use the hyperdistance, defined as the quadrature sum over the distances between all (six) pairs of nucleons i and j
\begin{equation}
\rho^2 = \frac{1}{4} \sum_{i<j} (\vec{x}_i-\vec{x}_j)^2, \label{eqn_rho}
\end{equation}
and to make a substitution to the standard wave function 
\begin{equation}
\chi (\rho) = \rho^4 \psi (\rho).
\end{equation}
The two wave functions are normalized as follows:
\begin{equation}
\int |\psi(\rho)|^2 \rho^8 d\rho = \int |\chi(\rho)|^2 d\rho = 1
\end{equation}
Fig. \ref{fig_ground} shows the probability distribution in hyperdistance of the ground state wave function. The inter-nucleon potential used reproduces the binding energy of $^4 He$ with $\langle E \rangle \simeq$ -28 MeV. This model also gives $\langle \rho \rangle =$ 2.24 fm; Assuming a tetrahedral configuration, one calculates a mean inter-nucleon distance $\langle r_{NN} \rangle =$ 1.83 fm. 

\section{Convergence and isotropy of the periodic box setup}

Perfect periodic boundary conditions can be imposed in numerical simulation only by an infinite number of boxes with image particles identical to the main simulation box. This, of course, would require infinite computational power and thus one must include only a small number of boxes. The question becomes then: what is the smallest number of boxes one should include to accurately include the effects of the periodic boundaries? The most important measure of this is the convergence of the desired observables - how the output valuables vary with the number of boxes. Here, we consider three configurations: a single periodic box with no images, a box with six images - one attached to each face of the box (7 total boxes), and a box enclosed by boxes touching every face, edge, and corner (27 total boxes).
\begin{figure}[h]
	\begin{centering}
	\includegraphics[width=0.45\linewidth]{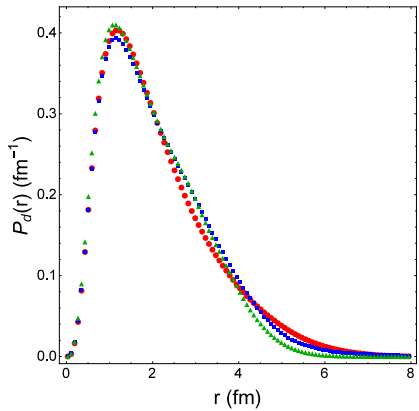}
	\caption{ Normalized probability distributions $P_d (r)$ of the deuteron system with 1 box (blue $\bullet$), 7 boxes (red $\blacksquare$), and 27 boxes (green $\blacktriangle$)}
	\end{centering}
\end{figure}
The most obvious consideration of the effects of the periodic boxes is the correction to the distributions of the system by the interactions of nucleons with images in other boxes, as the potential energy due to inter-box interactions affects the Metropolis updates. Additionally we look for anisotropy introduced by the finite number of images, which breaks spherical symmetry of the system. 
 
To test these properties of the setups, we consider the simplest system, the Walecka deuteron, in a box such that the nucleon density $n_N = 0.054$ fm$^{-3}$, the largest density considered in the main work. To test for anisotropy, the distribution of the nucleons' position $P(\theta,\phi)$, where $\theta$ and $\phi$ are the standard angles in spherical coordinates relative to the z-axis at the center of the box is measured and decomposed into real spherical harmonics
\begin{equation}
P(\theta ,\phi) = \sum_{\ell = 0}^{\infty} \sum_{m= -\ell}^{\ell} C_{\ell m} Y_{\ell m}(\theta ,\phi).
\end{equation}
In the case of a perfectly isotropic distribution $C_{\ell, m} = 0$ for all $\ell \ne 0$.
 
 We find, for our 1-box and 7-box setup, all coefficients of the expansion $C_{\ell, m} = 0$, except for $C_{2,0}$. The fact that the coefficient is nearly the same for both setups suggests that this is an artifact of the boundaries of the box itself rather than the images. The fact that nucleons are only moved to the opposite side of the box when their center of mass crosses the boundary may cause such effects. This may be due to the fact that the $Y_{2,0}$ has no probability along the 'diagonal' directions and the cubic geometry differentiates the directions along the diagonals and axes. Using a system with more images in the diagonal directions, such as the 27-box setup, may alleviate this anisotropy, but it would increase computation time and result in reduced statistics, so we do not do so. However, the main point of the comparison is to check that the 7-box setup used throughout this work does not introduce any additional anisotropy to the system. This seems to be the case as the coefficients are all equal within uncertainty in both setups.

\end{document}